\documentclass{webofc}
\usepackage[varg]{txfonts}   % Web of Conferences font
\usepackage{lineno}
\usepackage{hyperref}

\newcommand{\jpsi}{\ensuremath{\rm J/\psi}\xspace}

\newcommand{\s}{\ensuremath{\sqrt{s}}\xspace}
\newcommand{\snn}{\ensuremath{\sqrt{s_{\rm NN}}}\xspace}

\newcommand{\pt}{\ensuremath{p_{\text{T}}}\xspace}

\newcommand{\mee}{\ensuremath{m_{\text{ee}}}\xspace} 

\newcommand{\ccBar}{\ensuremath{c\bar{c}}\xspace}

\newcommand{\RAA}{\ensuremath{R_{\text{AA}}}\xspace}

\begin{document}
\title{Centrality and transverse momentum dependence of $J/\psi$ production in Pb-Pb collisions at $\sqrt{s_{\rm NN}}$  =  5.02 TeV at mid-rapidity with ALICE}

\author{\firstname{Dennis Weiser}\inst{1}\fnsep\thanks{\email{dennis.weiser@cern.ch}} \lastname{for the ALICE collaboration}  % etc.
}

\institute{Physikalisches Institut, Im Neuenheimer Feld 226, Heidelberg
          }

\abstract{%
  We present new results for the nuclear modification factor \RAA of \jpsi mesons as a function of centrality and transverse momentum at mid-rapidity. The measurement is carried out with the ALICE central barrel detectors in the acceptance range $|y|$ < 0.9 and \pt > 0 in the dielectron decay channel.
}
\maketitle
\section{Introduction}
\label{intro}
The measurement of \jpsi production is a key measurement to study deconfinement in nucleus-nucleus collisions. 
The initial idea was that the high color density in the medium prevents the 
formation of charmonia~\cite{matsui}. This would be observable as a suppression of the yield of \jpsi mesons in AA 
collisions compared to pp collisions scaled to binary nucleon-nucleon collisions. This suppression of the \jpsi 
yield in nucleus-nucleus collisions has been observed at lower energies at SPS and RHIC~\cite{na502005jpsiRaa, 
phenix2007JpsiRaaAtMidRapInAuAu200GeV, phenix2011JpsiRaaAtFwdRapInAuAu200GeV, phenix2008JpsiRaaInCuCu}. \\
At higher center-of-mass energies at the LHC the additional production of \jpsi mesons via the (re)combination  
of $\ccBar$ quarks is possible due to the increasing charm cross section $\sigma_{c\bar{c}}$. There are two main model 
scenarios describing the production of \jpsi by (re)combination: Thermal models and transport models. The thermal models assume 
a thermalization of charm in the medium and propose the production of \jpsi through statistical weights at the phase 
boundary \cite{stachel}. Transport models describe the production and dissociation of \jpsi as a dynamical 
process already before hadronization. Throughout the evolution of the medium partial destruction of \jpsi is 
counterbalanced by continuous production by (re)combination \cite{thews}.
The data from the LHC Run 1, taken at a collision energy of 2.76 TeV, have indeed shown that a (re)combination process is at work 
at this energy \cite{RAArun1}. However, the discussion on the details of this (re)combination process described by 
the different model scenarios still did not come to a conclusive picture.  
The newly taken data at \snn = 5.02 TeV in the LHC Run 2 provide further insights to the possible production mechanisms, due to more precise and differential measurements enabled by the higher available statistics.

\newpage
\section{Analysis}
\label{sec-1}
The analysis is based on a Pb-Pb dataset corresponding to an integrated luminosity of $\mathcal{L}_{\text{int}}$ $\approx$ 10 $\mu b^{-1}$ taken at \snn = 5.02 TeV with the ALICE detector \cite{ALICE}. \jpsi mesons are reconstructed at mid-rapidity ($|y| < 0.9$) and down to transverse momentum \pt = 0 in the dielectron decay channel. The Inner Tracking System (ITS) and the Time Projection Chamber (TPC) are used for track reconstruction. Secondary particles are rejected using the ITS. Electrons are identified using the specific energy loss (d$E$/d$x$) in the TPC. Combinatorial background introduced by electrons stemming from photon conversions is reduced by removing electrons which form pairs with very low invariant mass (\mee $<$ 50 MeV) from the pairing.\\
For the description of the background the event mixing technique is used. After subtraction of the background the signal is extracted in the invariant mass window 2.92 $<$ \mee $<$ 3.16 GeV/$c^{2}$ by bin counting. The loss of signal due to the tail stemming from bremsstrahlung is taken into account as part of the efficiency.\\
The \jpsi proton-proton reference cross-section at \s = 5.02 TeV is obtained from an interpolation procedure taking into account measurements at lower and higher collision energies \cite{ppphenix, ppcdf, ppalicehigh, ppalicelow}.

%For bibliography use \cite{RefJ}

%For one-column wide figures use syntax of figure~\ref{fig-1}
%\begin{figure}[h]
%\centering
%\includegraphics[width=0.4\textwidth,clip]{MInv_Prel/2017-Feb-02-MinvJpsi_EvSel_EM_standard_Cent0to10.pdf}
%\label{fig-1}       % Give a unique label
%\end{figure}

\section{Results}
\label{sec-2}
\begin{figure*}[h]
\centering
\includegraphics[width=0.45\textwidth,clip]{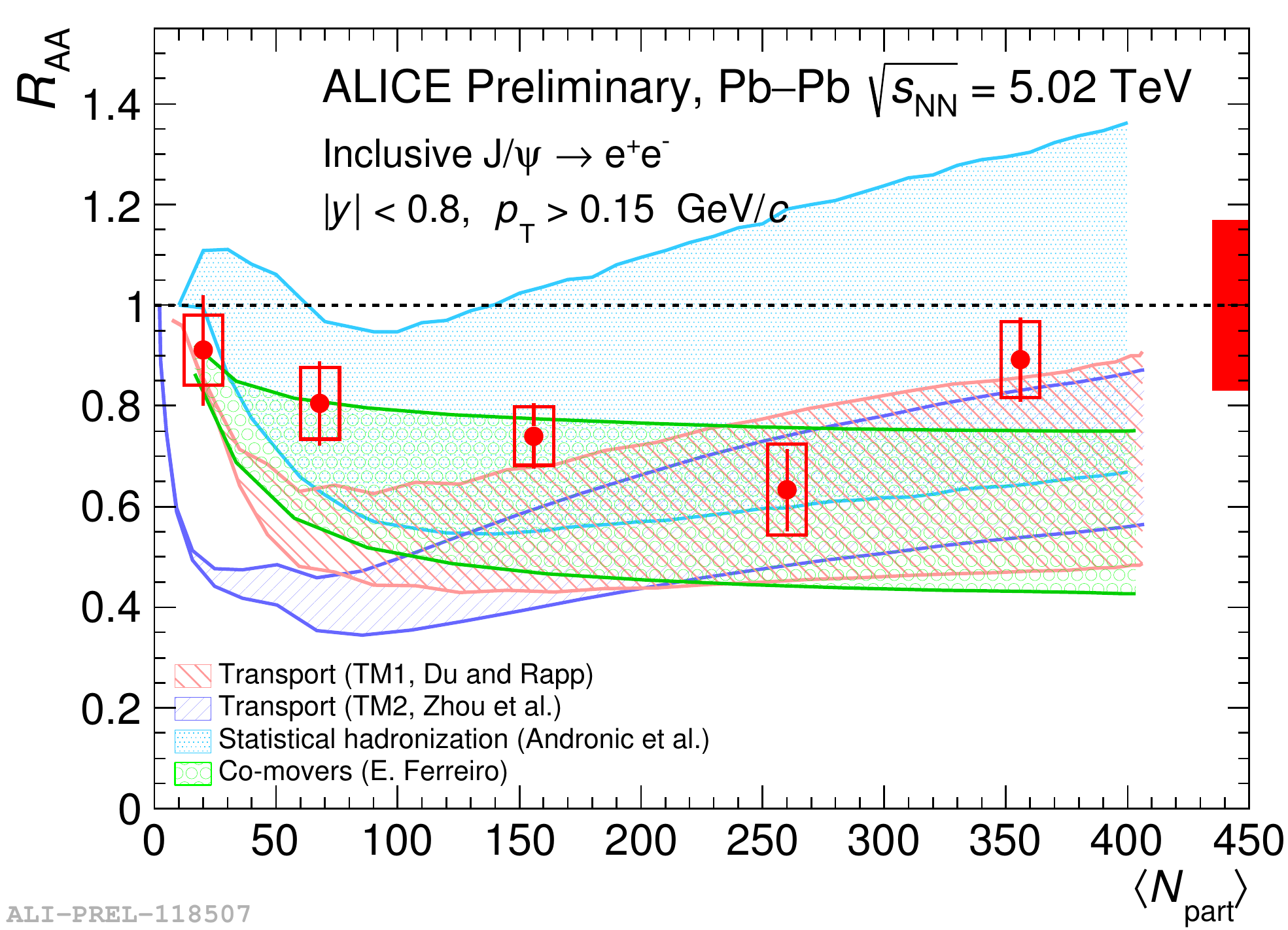}
\includegraphics[width=0.45\textwidth,clip]{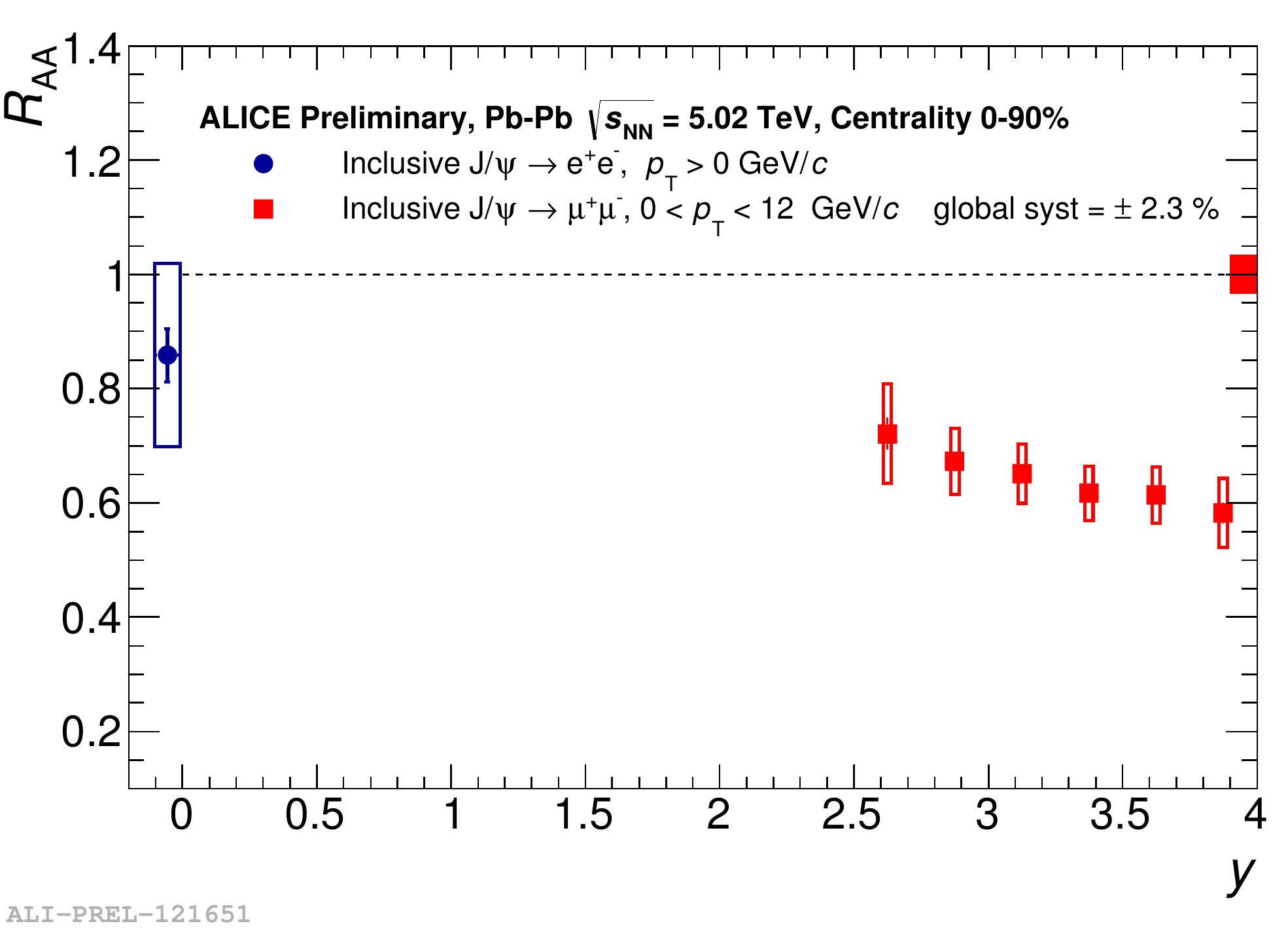}
%\vspace*{5cm}       % Give the correct figure height in cm
\caption{Centrality (left) and rapidity (right) dependent nuclear modification factor. The former is compared to predictions from a thermal model \cite{Andronic}, transport models \cite{Rapp11, Rapp15, Peng} and a comover model \cite{elena14, elena15}.}
\label{fig-2}       % Give a unique label
\end{figure*}

The nuclear modification factor \RAA for \jpsi mesons is obtained as a function of centrality, rapidity and transverse momentum. For the mid-rapidity measurement the \jpsi candidates with \pt $<$ 150 MeV have been excluded to avoid a contamination with photoproduced \jpsi's which mostly play a role in peripheral collisions. Figure \ref{fig-2} shows the nuclear modification factor as a function of the mean number of participants $\langle N_{\text{part}} \rangle$. The data points show no strong dependence on centrality; for most central collisions a hint of an increase can be noted which is also consistent with a fluctuation. The models contain a significant component of \jpsi production by (re)combination and describe the data rather well within their uncertainties. The model uncertainties are dominated by the imprecise knowledge of the total charm cross-section in pp collisions and the modification of the parton distribution functions in the Pb nucleus (shadowing). The global normalization uncertainty for the data points (red box) is dominated by the \jpsi pp reference cross section and amounts to 16.6\%.

%Don't forget to give each section, subsection, subsubsection, and
%paragraph a unique label (see Sect.~\ref{sec-1}).

%For two-column wide figures use syntax of figure~\ref{fig-2}

\begin{figure}[h]
\centering
\includegraphics[width=0.4\textwidth,clip]{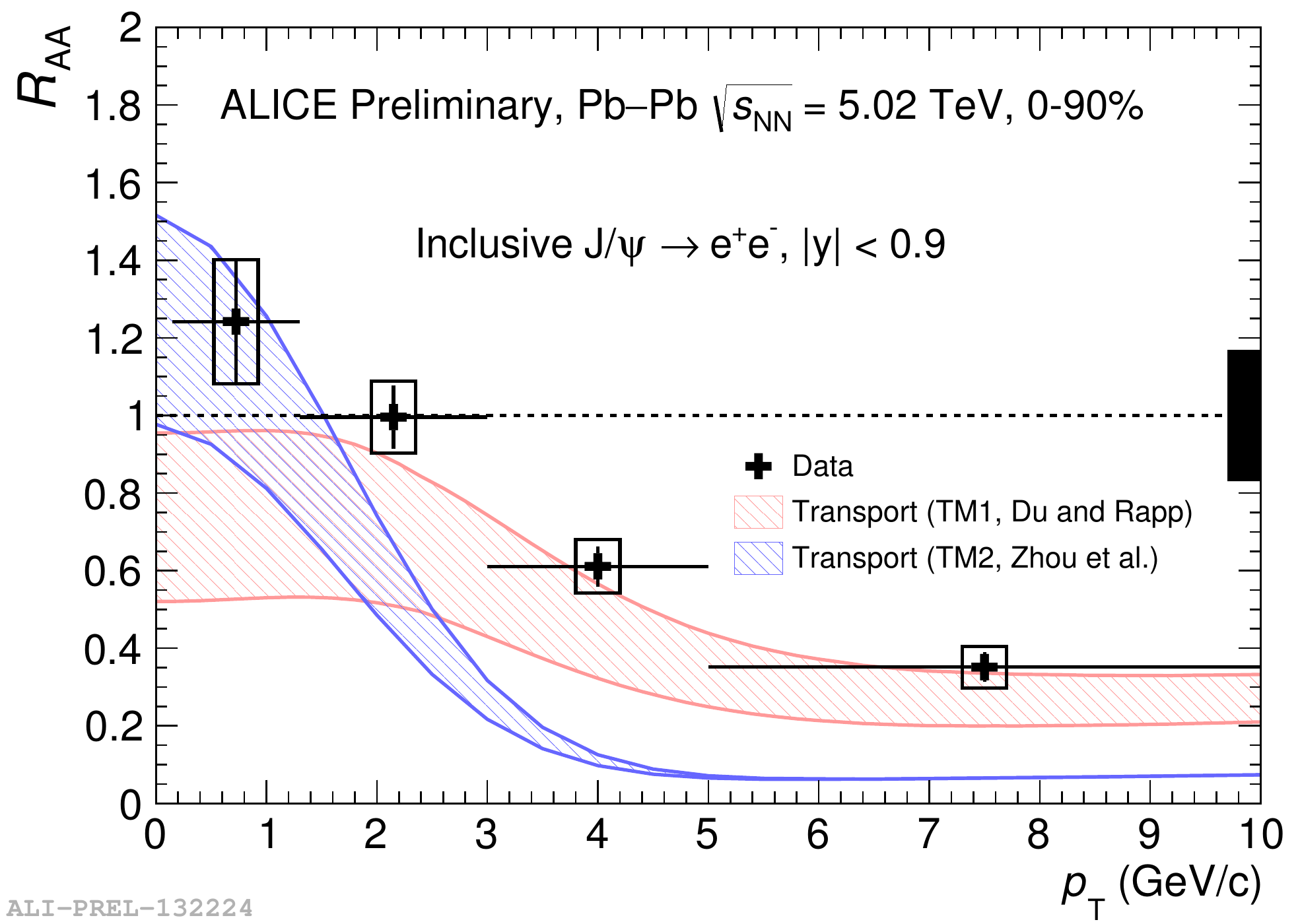}
\caption{The nuclear modification factor in the 0--90\% centrality range as function of \pt compared to two transport models \cite{Rapp11, Rapp15, Peng}.}
\label{fig-3}       % Give a unique label
\end{figure}

For the rapidity dependent measurement (Fig. \ref{fig-2} right panel) the normalization uncertainty for the mid-rapidity measurement is included in the mid-rapidity uncertainty box. The data show a trend of an increase in \RAA from forward to mid-rapidity. This is in line with the expectation from (re)combination models due to the increase of charm cross-section from forward to mid-rapidity. \\
The nuclear modification factor as function of \pt is shown in Fig. \ref{fig-3} for the full centrality range. The data show a stronger suppression with increasing \pt. This can be interpreted with the relevance of (re)combination at low \pt while at high \pt color screening or jet-quenching mechanisms should become dominant. The data are confronted with two transport model predictions. The model TM1 \cite{Rapp11, Rapp15} slightly undershoots the data, but can describe the data when all uncertainties are taken into account. The model TM2 \cite{Peng} clearly undershoots the data at high \pt. One might ask whether (re)combination is still relevant at high \pt or whether the hard production and/or B feed-down is modeled accurately enough in this model.

%For one-column wide figures use syntax of figure~\ref{fig-1}

\begin{figure*}[h]
\centering
\includegraphics[width=0.45\textwidth,clip]{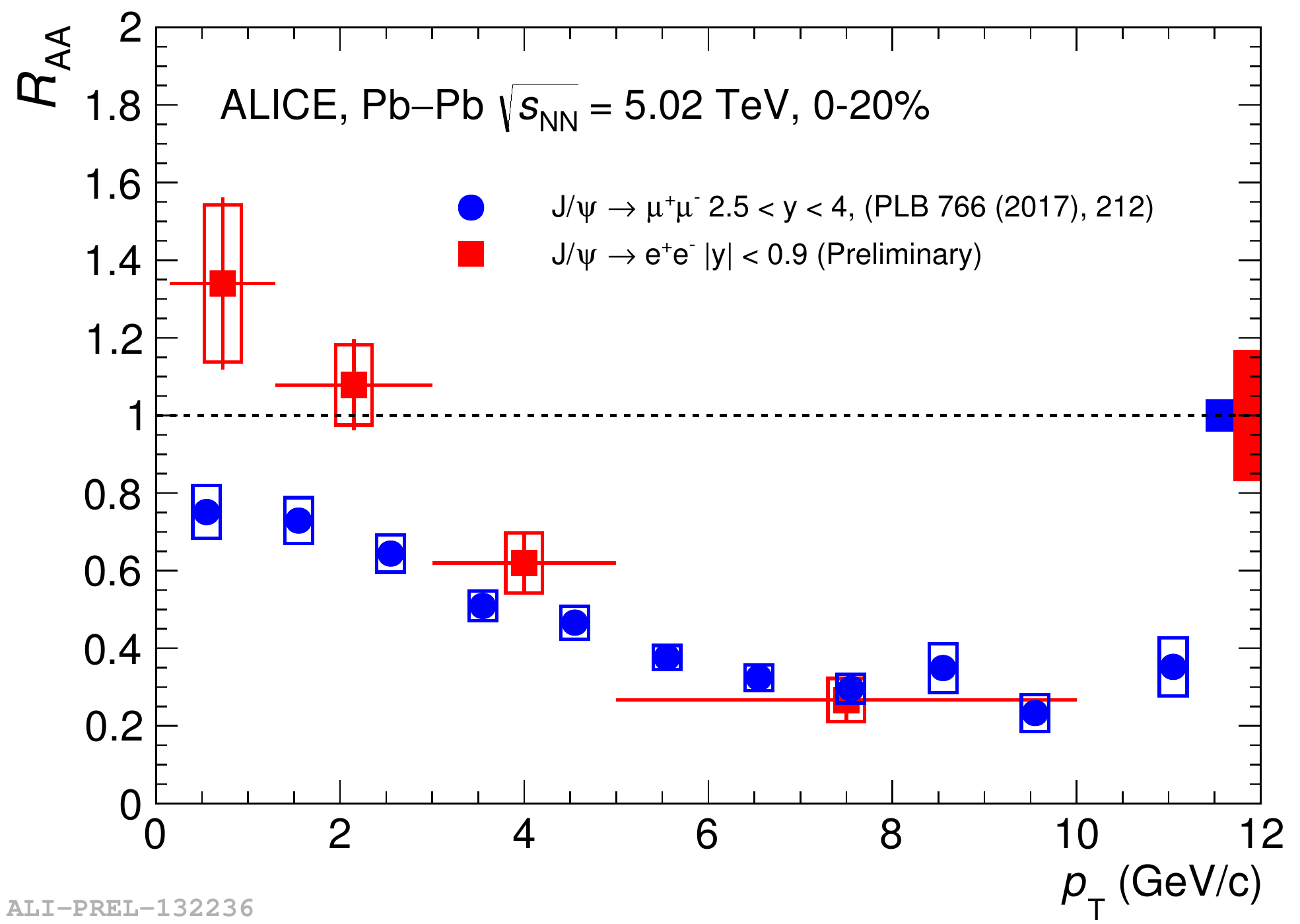}
\includegraphics[width=0.45\textwidth,clip]{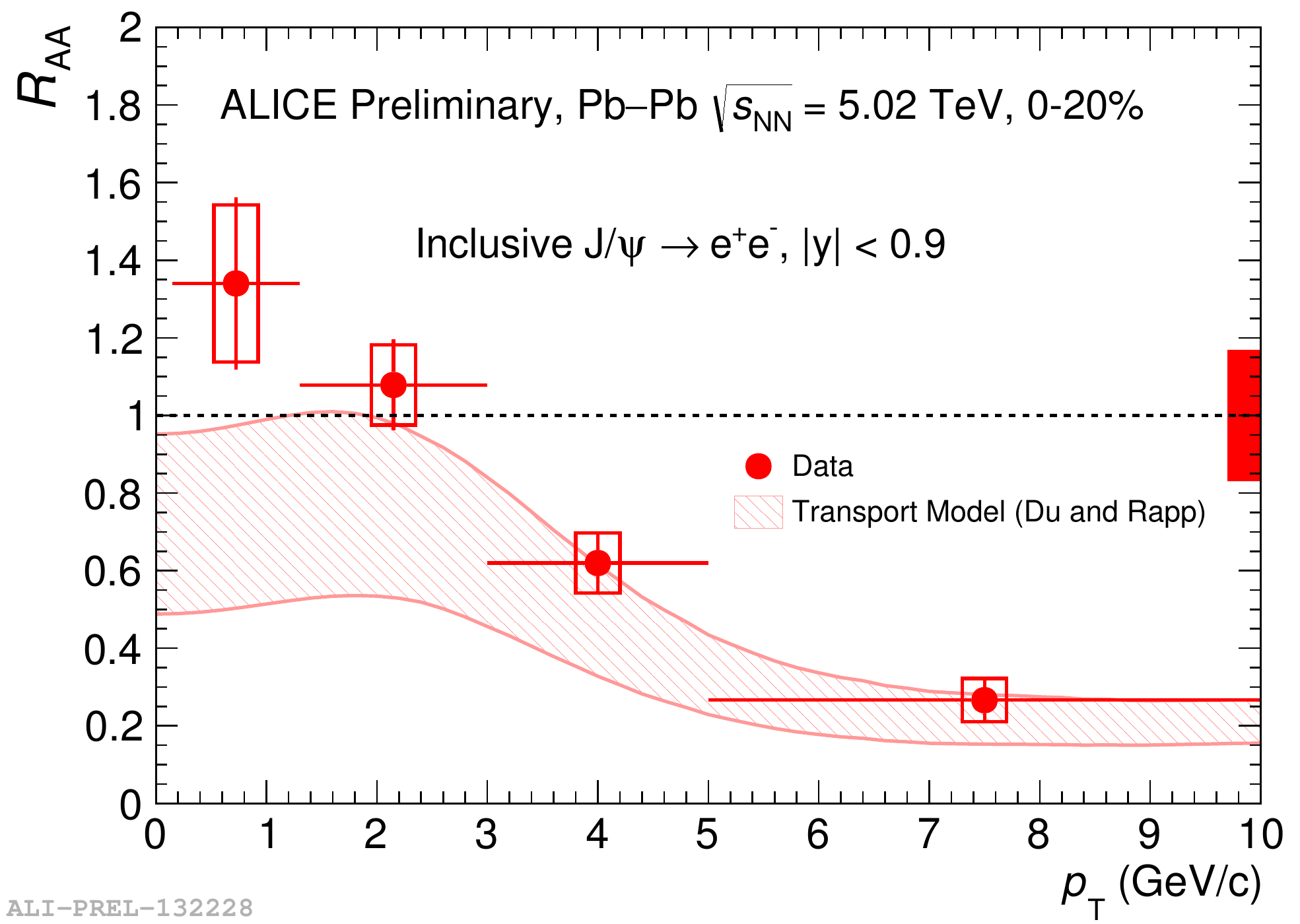}
%\vspace*{5cm}       % Give the correct figure height in cm
\caption{The nuclear modification factor measured in the 0--20\% centrality range compared to the forward measurement (left plot) and compared to a transport model prediction \cite{Rapp11, Rapp15} (right plot).}
\label{fig-4}       % Give a unique label
\end{figure*}

The transverse momentum dependent nuclear modification factor measured in the 0--20\% centrality range is shown in Fig. \ref{fig-4} compared to the forward measurement and compared to a transport model prediction. The general evolution of the \RAA with \pt is similar as for the centrality integrated case. Special emphasize should be given to the first two data points which exceed the muon measurement at low \pt. Again, the transport model (TM1) can describe the data when considering all uncertainties.

%For figure with sidecaption legend use syntax of figure
%\begin{figure}
%\centering
%\sidecaption
%\includegraphics[width=5cm,clip]{tiger}
%\label{fig-3}       % Give a unique label
%\end{figure}

\section{Conclusions}

The newly taken data by ALICE demonstrate that (re)combination plays an essential role for low \pt \jpsi production in Pb-Pb collisions at the highest available collision energies. At low \pt, a hint for a \jpsi nuclear modification factor exceeding unity has been observed. This observation is consistent with (re)combination models and thus supports the expectation of deconfinement and thermalization of charm in nucleus-nucleus collisions at LHC energies. Future measurements of the total charm cross-section in Pb-Pb collisions as planned by ALICE for the LHC Run 3 will help to reduce the model uncertainties and will thus allow the models to be put under close scrutiny when confronted with data.

% BibTeX or Biber users please use (the style is already called in the class, ensure that the "woc.bst" style is in your local directory)
% \bibliography{name or your bibliography database}

%

\end{document}